# A novel method for recommendation systems using invasive weed optimization

Fahimeh Soltaninejad[a], Amir Jalaly Bidgoly[a,*]

[a]Department of Computer Engineering, University of Qom, Qom, Iran

*Corresponding author.
E-mail addresses: Jalaly@qom.ac.ir

**Abstract**

One of the popular approaches in recommendation systems is Collaborative Filtering (CF). The most significant step in CF is choosing the appropriate set of users. For this purpose, similarity measures are usually used for computing the similarity between a specific user and the other users. This paper proposes a new invasive weed optimization (IWO) based CF approach that uses users' context to identify important and effective users set. By using a newly defined similarity measure based on both rating values and a measure values called confidence, the proposed approach calculates the similarity between users and thus identifies and filters the most similar users to a specific user. It then uses IWO to calculate the importance degree of users and finally, by using the identified important users and their importance degrees it predicts unknown ratings. To evaluate the proposed method, several experiments have been performed on two known real world datasets and the results show that the proposed method improves the state of the art results up to 15% in terms of Root Mean Square Error (RMSE) and Mean Absolute Error (MAE).

## 1. Introduction

Recommendation Systems (RSs) are useful tools which help users to find their favourite information in the great mass of information. Recommendation systems have been proposed as a solution to the information overload problem. These systems provide recommendations for users according to their interests and preferences. Recommendation systems have been used in various fields such as recommending book [1], news [2, 3], video [4], music [5] and items [6, 7].

The two most important approaches in recommendation systems are Content based Filtering (CB) and Collaborative Filtering (CF) [8, 9]. In CB approaches, recommendations are generated according to the interests of a specific user, without looking at the interests of other users. In fact, in these approaches, the content of the items is considered and based on the characteristics of the user's previous favourite items, items with similar characteristics are recommended to him. CF approaches use the interests of other users to generate recommendations for a specific user. In fact, in these approaches, users similar to a specific user are identified and then their interests are used to generate recommendations. The first and most significant step in CF is finding an appropriate set of users. Different methods are used for this purpose. In order to compute the similarity between users, various similarity measures including Euclidean distance, cosine similarity, Pearson correlation coefficient (PCC) are used.

In recent years, the use of meta heuristic methods in the field of recommendation systems has received much attention. Meta heuristic methods are nature based approaches which solve various complicated optimization problems[10]. Experiments have shown that the use of meta heuristic methods in recommendation systems has led to improved recommendation results. The meta heuristic algorithms used in the recommendation system approaches include Genetic Algorithm (GA), Particle Swarm Optimization (PSO), Ant Colony Optimization (ACO) and cuckoo search.

Inspired by [11] and [12], this paper proposes a new *Invasive Weed Optimization* algorithm based CF approach that uses users' context and filters similar users based on a measure called *confidence*. The proposed method consists of three primary phases: selecting and filtering important users, calculating the importance of selected users and predicting unknown ratings. In the first phase, the most similar users to a specific user are identified and selected as important users using a newly defined similarity measure based on a measure called confidence. In the second phase, the importance of the selected users is determined by using the IWO algorithm. In the third phase, specific user's unknown ratings of items are estimated using the selected important users and their importance degrees. To evaluate the proposed method, various experiments have been done on two popular real world datasets, and the results show that the proposed method has well performance and improves the results of the state of the art methods up to 15% in terms of Root Mean Square Error (RMSE) and Mean Absolute Error (MAE).

The rest of the paper is as follows. In Section 2, the related works and background are described. The proposed method is described in Section 3. In Section 4, details of the performed experiments are reported including datasets, the evaluation measures, and the achieved results. Finally, the paper ends in Section 5 with conclusion and the future directions.

## 2. Related works and background

In this section, previous researches that have been done in the field of recommendation systems, specifically approaches which are based on meta heuristic methods are reviewed. In the continuation of this section, the IWO algorithm and its basic concepts are described.

### 2.1 Metaheuristic methods based recommendation system approaches

Recently, meta heuristic methods based recommendation systems have attracted a lot of attention [11-25]. Metaheuristic algorithms are nature inspired algorithms which find the best solution among all possible solutions. In other words, these algorithms find the optimal or near optimal solution by searching a search space. The use of meta heuristic methods in recommendation systems has improved the results of the recommendations. Recommendation system approaches have been proposed based on meta heuristic methods including Genetic Algorithm (GA), Particle Swarm Optimization (PSO), Ant Colony Optimization (ACO) and Cuckoo Search. For example, Fong et al. [13] have proposed a GA based approach to combine CF and CB methods. The proposed approach uses GA to weight features of the extracted feature vector per user.

In [20], Gao et al. have proposed a model that combines the prediction results of different recommendation approaches using GA in order to achieve the best results. In the proposed model, a weight vector is considered for each recommender, which shows the ability of that recommender for predicting the rating of items. The weight vectors are optimized by

GA so that the best ratings for the items are predicted. AI-Shamri et al. [23] have proposed a user model based on user characteristics set including both ratings of a user for items and the content of items, which can be used to find similar users using GA. In fact, the proposed approach uses GA to find the appropriate weight per user model characteristic and then uses a fuzzy distance measure to find similar users. Bobadilla et al. [19] have introduced a CF based approach which utilizes a new similarity measure for identifying the most similar users to a specific user. This approach uses GA to find the optimal similarity function. It should be noted that this proposed approach only uses user's rating information.

By using GA and a multidimensional information model, Salehi et al. [21] have proposed a recommendation system for learning materials according to their implicit or explicit attributes. In this work, the weights of implicit attributes of learning materials are optimized using GA. In [22], Yilmaz et al. use GA to improve the calculated similarity values between a specific user and the other users and then use these values in the process of predicting unknown ratings. Da Silva et al. [24] also have combined the results of different CF based recommendation approaches using GA to produce the best results.

Bedi and sharma have proposed TARS [11], a recommender system based on ACO and trust relationships between users. TARS employs ACO in order to identify trustworthy users of a specific user and Then generates recommendations using these identified users. Liu et al. [18] have developed an ACO based approach which considers people behaviour in search of favourite information as ants behaviour in search of food, and therefore by simulating ants searching behaviour, it generates efficient recommendations and discovers knowledge. The proposed method also uses semantic annotation to speed up the search process. In the work of Gohari et al. [14], A trust based recommendation system using ACO has been proposed, which also considers context dependency in its trust computation. It uses ACO to find the best trust paths in the trust network and select outstanding neighbours of a specific user. Parvin et al. [12] have proposed TCFACO, a trust based CF approach which employs ACO to find similarity weights of users.

PSO algorithm has been employed in the work of Ujjin and Bentley [15], in order to learn users preferences and generate appropriate recommendations. In the work of Wasid and Kant [17], a PSO approach for CF based recommendation system through fuzzy features has been proposed. In this work, the PSO algorithm has been used in order to weight different individual features of users. also, fuzzy sets have been used to represent user features efficiently. In the work of Choudhary et al. [16], a PSO based approach for a multi criteria recommendation system has been proposed using effective similarity measures. In fact, in this work, different criteria have been considered and based on each criterion, similarity has been calculated using an effective similarity measure. Then the results of similarity calculation over different criteria have been aggregated with the help of PSO algorithm which weights different criteria according to different users and thus a multi criteria recommender system has been built. Katarya and Verma [25] have proposed a recommendation system in the field of movie, that uses k-means clustering method and cuckoo search algorithm. The proposed approach uses k-mean clustering in order to categorize users and the cuckoo search algorithm in order to optimize categories. Finally, it generates recommendations using obtained similar users of a specific user.

In summary, all of the above methods demonstrate the importance of employing metaheuristic algorithms in the recommendation process and especially in weighting users. The proposed method in this paper uses the IWO random process to find neighbouring users with a measure called confidence, which provides better convergence and accuracy results.

## 2.2 Invasive Weed Optimization algorithm(IWO)

Mehrabian and Lucas [26] proposed invasive weed optimization algorithm (IWO) in 2006. This algorithm is inspired by nature and can be utilized in order to find solution for various optimization problems in the science and engineering fields. The origin of IWO algorithm is the method of reproduction and survival of weeds, and it is a process that is both optimal and resistant. One of the important factors in weeds growth is R/K selection theory. In the theory of choice R the strategy is quantity increasing but in the theory of choice K the strategy is quality increasing (i.e. quality improvement). In the IWO algorithm, movement takes place from the R-selection rule to the K-selection rule (i.e. from trying all places to trying only places that are optimal). The goal of the IWO is finding the best place to live. The steps of IWO are as follows:

1. In the first step, a population of initial solutions (weeds) is produced.
2. In the second step, the weed is allowed to reproduce, based on the fitness value of each solution. The number of child seeds is determined according to the following equation:

$$s = \left\lfloor s_{min} + (s_{max} - s_{min})\left(\frac{f - f_{worst}}{f_{best} - f_{worst}}\right) \right\rfloor \tag{1}$$

3. In the third step, the child seeds are scattered around the mother weed using a normal distribution as follows. If the total number of weeds reaches a predefined maximum value, all of them are sorted and the worse ones are removed.

$$\Delta x_i \sim N(0, \sigma_t^2) \quad \sigma_t = \left(\frac{T-t}{T}\right)^n (\sigma_{initial} - \sigma_{final}) + \sigma_{final} \tag{2}$$

4. In the fourth step, if the stop conditions are not met, repetition is done from step (2), otherwise the algorithm ends.

Fig. 1 shows the pseudo code of the IWO.

```
Generate a random population of solutions(W);
For iter=1 to the maximum number of generations(MaxIt)
Evaluate the objective function for each individual in population;
Compute maximum and minimum fitness in the colony;
For each individual in population
    iv.   Compute the number of seeds of each individual, corresponding to its fitness;
    v.    Randomly distribute the generated seeds over the search space with normal distribution
          around the parent weed;
    vi.   Add generated seeds to the solution set;
If([W]=N)>P_max
    iii.  Sort the Population in descending order of their fitness;
    iv.   Truncate population of weeds with smaller fitness until N=P_max;
Next iteration.
```

*Figure1: The Pseudo code of the IWO*

## 3. Proposed Method

The proposed method consists of three primary phases: selecting and filtering important users, calculating the importance of selected users (i.e. weighting them) and predicting unknown ratings. In the first phase, similarity values between a specific user and the other users are computed according to a measure called *confidence* and users that their corresponding similarity are above a threshold picked as the most similar users to specific user and called important users. In this phase, *Pearson correlation coefficient* (PCC) is used in combination with *confidence* measure in order to filter important and effective users. In the second phase, the importance degree of the selected users is determined using IWO. In the last phase, specific user's unknown ratings of items are estimated using the selected important users and their importance degrees. A summary of the proposed method phases is shown in Fig.2. Further details of each phase are explained in the following subsections.

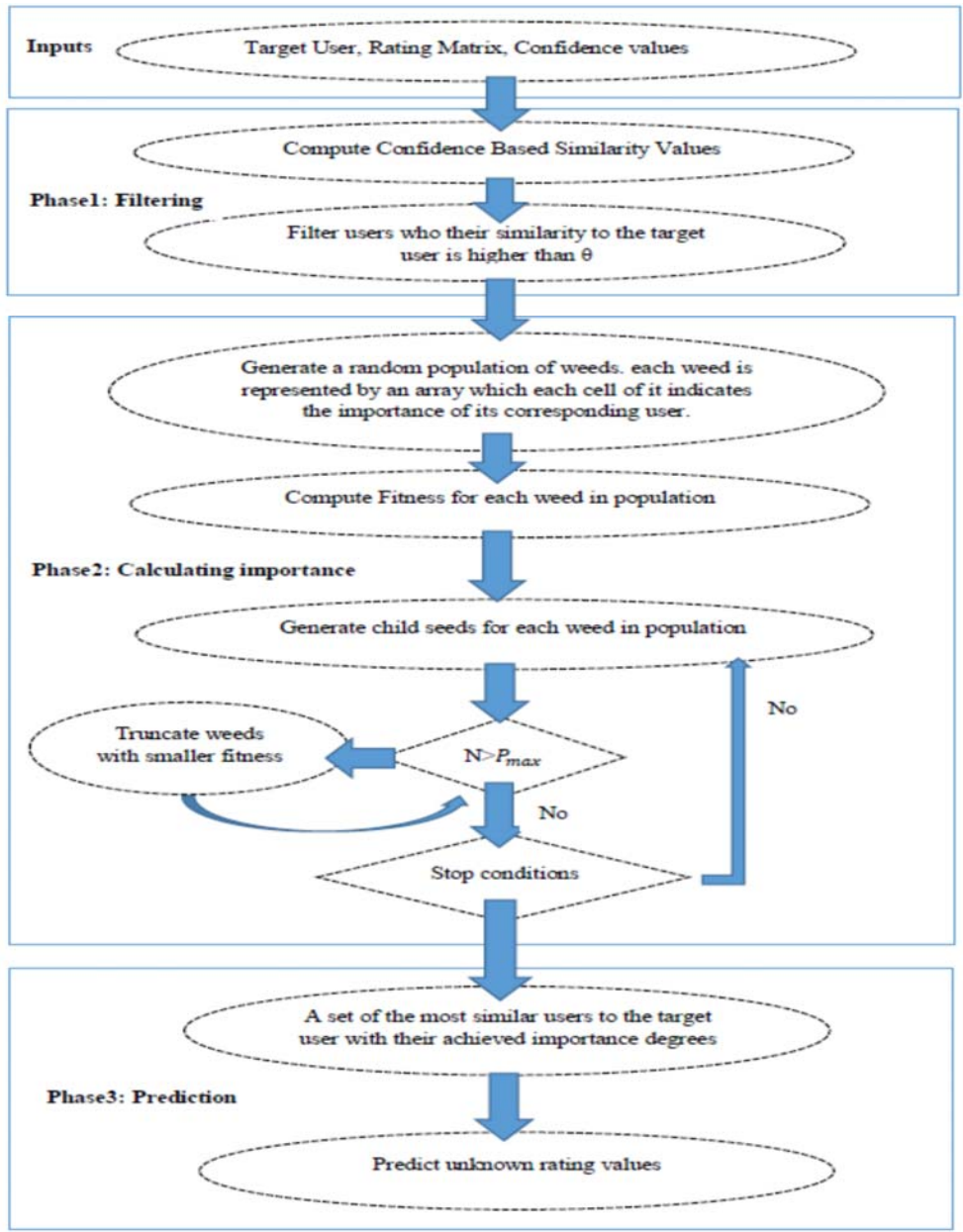

*Figure 2: A summary of the Proposed Model Phases*

## 3.1 Filtering important users

The purpose of this phase is to compute the similarity values between a specific user and the other users according to the rating values and proposed confidence measure. Using the confidence measure along with the similarity measures, not only enables the method to measure the similarity of any users but also helps to consider the reliability of this similarity in the final estimation. Suppose $u$ and $v$ are two users of the system, the estimated similarity between these users are denoted by $W_{u,v}$ and defined as follows:

$$W_{u,v} = \begin{cases} \frac{2*sim(u,v)*conf(v|u)}{sim(u,v)+conf(v|u)} & if\ sim(u,v)!=0\ and\ conf(v|u)!=0 \\ k*conf(v|u) & if\ sim(u,v)==0\ and\ conf(v|u)!=0 \\ 0 & if\ sim(u,v)==0\ and\ conf(v|u)==0 \end{cases} \quad (3)$$

where $sim(u,v)$ and $conf(v|u)$ are the naive similarity measure between these users and the confidence about this measure respectively. Also, $k$ is a small constant value. In general, there are several measures for estimating similarity including the *Pearson correlation coefficient, cosine similarity* and *Euclidean distance*. The proposed method in this paper uses *Pearson correlation coefficient* as follows to calculate the similarity between user $u$ and user $v$ which can achieve the best results between these methods:

$$sim(u,v) = \begin{cases} P_{u,v} = \frac{\sum_{i \in A_{u,v}}(r_{u,i}-\overline{r_u})(r_{v,i}-\overline{r_v})}{\sqrt{\sum_{i \in A_{u,v}}(r_{u,i}-\overline{r_u})^2}\sqrt{\sum_{i \in A_{u,v}}(r_{v,i}-\overline{r_v})^2}} & if\ P_{u,v} > 0 \\ 0 & otherwise, \end{cases} \quad (4)$$

where $A_{u,v}$ is a set of items rated by both user $u$ and user $v$, $r_{u,i}$ and $r_{v,i}$ respectively denote the rating values of user $u$ and user $v$ for an item $i$, and $\overline{r_u}$ and $\overline{r_v}$ are the mean of rating values of user $u$ and user $v$. Pearson correlation coefficient, denoted by $P_{u,v}$, is calculated on the ratings of items which rated by both users, in other words, on the ratings of items in $A_{u,v}$ for each user $u$ and user $v$. The value of $P_{u,v}$ lies in the range of -1 to 1 which the more values indicate the more similarity between the users. Since $P_{u,v} \leq 0$ indicates that $u$ and $v$ are not correlated, consequently zero and negative correlations are not considered and all those outputs are considered zero.

As mentioned before, the proposed confidence measure, denoted by $conf(v|u)$, helps to include the reliability in the similarity measurement. This measure is defined as follows:

$$conf(v|u) = \frac{no.\ of\ items\ rated\ by\ u\ and\ v\ in\ common + 1}{no.\ of\ items\ rated\ by\ u + 2} \quad (5)$$

Above equation indicates that in a reliable estimation, user $v$ should have rated most of the items that user $u$ has rated, otherwise the estimated similarity may not be very reliable. As seen in the above equation, the more rating overlap between user $u$ and user $v$ leads to the higher values of $conf(v|u)$. Finally, users who have a similarity value ($W_{u,v}$) above a defined threshold θ are identified as the neighbours of specific user (i.e. as important users). The set of important users selected for usage in the next phases is called $IU_u$ and is defined as follows:

$$IU_u = \{v \in U\ |\ W_{u,v} > \theta\} \quad (6)$$

where $U$ and $\theta$ represent the users set and the threshold of selecting important users respectively.

### 3.2 Calculating the importance of selected users using IWO

In this phase, importance degrees of the selected important users are determined using IWO algorithm. For this purpose, each solution is represented by a vector of users weights, denoted by $w$, where $w_v$ indicates the importance degree of selected user $v \in IU_u$ in the range of $[0,1]$. By using IWO, first a random population of solutions is generated and Then fitness is calculated for each solution. Depending on the fitness value, each solution is allowed to reproduce, and thus other solutions are produced around the mother solution. This process is repeated until the defined stop conditions are met, for example a certain fitness value is achieved. Finally, vector of optimal weights for the important users is obtained and therefore the importance degrees of them is calculated. The selected fitness function in this approach is *Mean Absolute Error* (MAE) which by using it, the quality of each solution is evaluated. MAE is a measure for indicating the deviation of the estimated ratings from the actual ones and is computed as follows:

$$Fitness(u) = \frac{\sum_{i=1}^{I_u}|\widehat{r_{u,i}}-r_{u,i}|}{|I_u|} \tag{7}$$

where $I_u$ shows the set of items which recommender system is supposed to predict ratings of them.

Finally, after finishing the search operation and reaching a suitable fitness value, the set of selected important users along with their importance degrees is achieved.

### 3.3 Predicting unknown ratings

In this phase, specific user's unknown ratings of items are estimated as follows using the selected important users and their obtained importance degrees:

$$\widehat{r_{u,i}} = \frac{\sum_{v \in IU_u} w_v r_{v,i}}{\sum_{v \in IU_u} w_v} \tag{8}$$

where $w_v$ and $r_{v,i}$ represent the importance degree of user $v$ and the actual rating of him on the item $i$ respectively. $IU_u$ is the set of selected important users and $\widehat{r_{u,i}}$ is the estimated rating of specific user $u$ on the item $i$.

### 4. Experiments and Results

Several experiments have been conducted on two popular datasets in order to evaluate the performance of the proposed approach. The experiments were done on a PC with a Core i7 processor and 8GB RAM, and the proposed method was implemented with the Python programming language. Further details including datasets, the evaluation measures, the parameter settings and the achieved results are explained in the following subsections.

### 4.1 Datasets

*Filmtrust*[1] and *Epinions*[2] datasets have been used in the experiments. *Filmtrust* is collected from a website where users can express their opinions about various movies. User ratings for movies and their reviews are available on this website. Users can rate different movies from 0.5 to 4. This dataset consists of rating information of 1508 users on 2071 items. *Epinions* is collected from epinions[3] website, a popular but currently closed website that has provided a platform for users to review and score various products. Users can rate different items from 1 to 5. This dataset consists of rating information of 40163 users on 139738 items.

### 4.2 Evaluation measures

Results are reported based on the two well-known measures named *Mean Absolute Error (*MAE) and *Root Mean Square Error* (RMSE). These two measures show the predictive accuracy of recommendation systems. Using MAE and RMSE, it is possible to assess how much the recommender system has been able to predict the user's actual ratings. In other words, the lower MAE and RMSE, the more accurate the recommender system predicts user ratings.

MAE is the average of all differences of the absolute value between the actual and the estimated rating. RMSE is the square root of the average of all the squared differences between the actual and the estimated rating. The calculation formula for MAE and RMSE is as follows.

$$MAE = \frac{1}{z}\sum_{(u,j)} |\widehat{r_{u,j}} - r_{u,j}| \qquad (9)$$

$$RMSE = \sqrt{\frac{1}{z}\sum_{(u,j)} (\widehat{r_{u,j}} - r_{u,j})^2} \qquad (10)$$

where $\widehat{r_{u,j}}$ and $r_{u,j}$ represent the estimated and the actual rating of user $u$ on the item $j$ respectively. Z represents the number of ratings which the recommender system predicts.

### 4.3 Parameter Settings

There are several adjustable parameters in the proposed method. In order to specify the values of these parameters, various experiments have been done. In these experiments, different values of these parameters have been examined and finally the following values have been achieved as the optimal values of the parameters: $s_{min}$=0, $s_{max}$=7, $\sigma_{initial}$=1, $\sigma_{final}$=0.001, n=5 and T(number of iterations) = 300. The number of the initial population and the final population in IWO algorithm are set to 10 and 200 respectively. Parameter $k$ in the similarity calculation equation is set to 0.2. Experiments showed that increasing the value of $\theta$, which is the threshold of selecting important users in the first phase, leads to better recommendation results. Consequently, the optimal value of 0.6 is achieved for $\theta$.

### 4.4 Results

In order to evaluate the proposed approach, various experiments have been done. The proposed approach has been compared with several known metaheuristic methods based collaborative filtering approaches including Babadilla [19], Yilmaz [22], TARS [11] and TCFACO [12].

---

[1] htttps://www.librec.net/datasets.html
[2] http://www.trustlet.org/epinions.html
[3] http://www.Epinions.com

Bobadilla recommender system uses GA in order to find the optimal similarity function according to rating information [19]. Yilmaz recommender system uses GA to improve the calculated similarity values between a specific user and the other users, and then uses those values in the process of predicting unknown ratings [22]. TARS is a trust based approach which uses ACO to identify trustworthy users of a specific user and then generates recommendations using those users [11]. TCFACO is another trust based CF approach that uses ACO to find similarity weights of users [12].

Tables 1 and 2 provide comparative results on *FilmTrust* and *Epinions* datasets, respectively. Compared to the other approaches, the proposed approach achieves less MAE and RMSE for both datasets and it improves results of the other approaches up to 15%. Better performance is due to the fact that the proposed approach is more accurate than the other approaches in filtering important and effective users. It calculates the importance degree of users using IWO algorithm and then uses those values to predict unknown ratings. The results show that the usage of importance values of users in the prediction phase has a great impact on improving the recommendation results.

*Table 1: Experiment results on Filmtrust dataset*

|  | MAE | RMSE |
|---|---|---|
| **Bobadilla** | 0.771 | 0.982 |
| **Yilmaz** | 0.685 | 0.912 |
| **TARS** | 0.662 | 0.872 |
| **TCFACO** | 0.561 | 0.764 |
| **Proposed Method** | **0.545** | **0.656** |

*Table 2: Experiment results on Epinions dataset*

|  | MAE | RMSE |
|---|---|---|
| **Bobadilla** | 0.862 | 1.124 |
| **Yilmaz** | 0.852 | 1.101 |
| **TARS** | 0.830 | 1.092 |
| **TCFACO** | 0.795 | 1.043 |
| **Proposed Method** | **0.710** | **0.961** |

## 5. Conclusions and Future directions

This paper proposed a new invasive weed optimization (IWO) based collaborative filtering approach for recommender systems. The proposed approach consists of three primary phases. In the first phase, a new similarity measure based on both rating values and defined measure values named confidence is proposed, which computes the similarity between a specific user and the other users. Next, users who have a similarity value above a defined threshold are identified and filtered as important users. In the second phase, the importance of the identified important users is calculated using the IWO algorithm. In the last phase, using the identified important users and their importance degrees, unknown ratings are predicted. To evaluate the proposed approach, various experiments has been done on Filmtrust and Epinions datasets. Results show that the proposed approach improves the state of the art results up to 15% in terms of Root Mean Square Error (RMSE) and Mean Absolute Error (MAE).

Future work to improve the proposed method is that in addition to the rating values and confidence measure values, extra information sources be considered in the first phase of the proposed method. In other words, other additional information such as social relationships of users be used in the process of identifying and filtering important users, and thereby the accuracy of the recommendation generation system be increased.

**References**

[1]     E. Rich, "User modelling via stereotypes," *Cognitive Science Journal,* vol. 3, no. 4, pp. 329–354, (1979).
[2]     J. A. Konstan, B. N. Miller, D. Maltz, J. L. Herlocker, L. R. Gordon, and J. Riedl, "GroupLens: Applying collaborative filtering to UseNet news," *Communications of the ACM,* vol. 40, no. 3, pp. 77–87, (1997).
[3]     P. Resnick, N. Lakovou, M. Sushak, P. Bergstrom, and J. Riedl, "GroupLens: An open architecture for collaborative filtering of Netnews," in *Proceedings of the 1994 ACM conference on Computer supported cooperative work*, (1994), pp. 175–186.
[4]     W. Hill, L. Stead, M. Rosenstein, and G. Furnas, "Recommending and evaluating choices in a virtual community of use," in *Proceedings of the SIGCHI conference on Human factors in computing systems*, (1995), pp. 194–201.
[5]     U. Shardanand and P. Maes, "Social information filtering: Algorithms for automating 'Word of Mouth'," in *Proceedings of the SIGCHI Conference on Human Factors in Computing Systems*, (1995), pp. 210-217.
[6]     G. Linden, B. Smith, and J. York, "Amazon. com recommendations: Item-to-item collaborative filtering," *IEEE Internet computing,* vol. 7, no. 1, pp. 76-80, (2003).
[7]     B. M. Sarwar, G. Karypis, J. A. Konstan, and J. Riedl, "Item-based collaborative filtering recommendation algorithms," in *Proceedings of the 10th international conference on World Wide Web*, (2001), pp. 285-295.
[8]     D. M. Pennock, E. Horvitz, S. Lawrence, and C. L. Giles, "Collaborative filtering by personality diagnosis: A hybrid memory-and model-based approach," in *Proceedings of the Sixteenth conference on Uncertainty in artificial intelligence* (2000), pp. 473-480.
[9]     J. Salter and N. Antonopoulos, "Cinema Screen recommender agent: combining collaborative and content based filtering," *IEEE Intelligent Systems,* vol. 21, no. 1, pp. 35-41, (2006).
[10]    M. Dorigo and L. M. Gambardella, "Ant colony system: a cooperative learning approach to the traveling salesman problem," *IEEE Transactions on evolutionary computation,* vol. 1, no. 1, pp. 53-66, (1997).
[11]    P. Bedi and R. Sharma, "Trust based recommender system using ant colony for trust computation," *Expert Systems with Applications,* vol. 39, no. 1, pp. 1183-1190, (2012).
[12]    H. Parvin, P. Moradi, and S. Esmaeili, "TCFACO: Trust-aware collaborative Filtering method based on ant colony optimization," *Expert Systems with Applications,* vol. 118, pp. 152-168, (2018).
[13]    S. Fong, Y. Ho, and Y. Hang, "Using genetic algorithm for hybrid modes of collaborative filtering in online recommenders," presented at the Eighth International Conference on Hybrid Intelligent Systems, (2008).
[14]    F. S. Gohari, H. Haghighi, and F. S. Aliee, " A semantic-enhanced trust based recommender system using ant colony optimization," *Applied Intelligence,* vol. 46, pp. 328-364, (2017).
[15]    S. Ujjin and P. J. Bentley, "Particle Swarm Optimization recommender system," in *Proceedings of the 2003 IEEE Swarm Intelligence Symposium*, (2003), pp. 124–131: IEEE.
[16]    P. Choudhary, V. Kant, and P. Dwivedi, "A Particle Swarm Optimization approach to multi criteria recommender system utilizing effective similarity measures," in *Proceedings of the 9th International Conference on Machine Learning and Computing*, (2017), pp. 81–85.
[17]    M. Wasid and V. Kant, "A particle swarm approach to collaborative filtering based recommender systems through fuzzy features," *Procedia Computer Science,* vol. 54, pp. 440-448, (2015).

[18]   M. Liu, S. Liu, X. Wang, M. Qu, and C. Hu, "Knowledge-domain semantic searching and recommendation based on improved ant colony algorithm," *Journal of Bionic Engineering,* vol. 10, pp. 532-540, (2013).

[19]   J. Bobadilla, F. Ortega, A. Hernando, and J. Alcalá, "Improving collaborative filtering recommender system results and performance using genetic algorithms," *Knowledge-Based Systems,* vol. 24, no. 8, pp. 1310-1316, (2011).

[20]   L. Gao and C. Li, "Hybrid personalized recommended model based on genetic algorithm," presented at the 4th International Conference on Wireless Communications, Networking and Mobile Computing ((2008).

[21]   M. Salehi, M. Pourzaferani, and S. A. Razavi, "Hybrid attribute-based recommender system for learning material using genetic algorithm and a multidimensional information model," *Egyptian Informatics Journal,* vol. 14, no. 1, pp. 67-78, (2013).

[22]   Y. Ar and E. Bostanci, "A genetic algorithm solution to the collaborative filtering problem," *Expert Systems with Applications,* vol. 61, pp. 122-128, (2016).

[23]   M. Y. H. Al-Shamri and K. K. Bharadwaj, "Fuzzy-genetic approach to recommender systems based on a novel hybrid user model," *Expert Systems with Applications,* vol. 35, no. 3, pp. 1386-1399, (2008).

[24]   E. Q. da Silva, C. G. Camilo-Junior, L. M. L. Pascoal, and T. C. Rosa, "An evolutionary approach for combining results of recommender systems techniques based on collaborative filtering," *Expert Systems with Applications,* vol. 53, pp. 204-218, (2016).

[25]   R. Katarya and O. P. Verma, "s," *Egyptian Informatics Journal,* vol. 18, no. 2, pp. 105-112, (2017).

[26]   A. R. Mehrabian and C. Lucas, " A novel numerical optimization algorithm inspired from weed colonization " *Ecological Informatics,* vol. 1, no. 4, pp. 355 – 366, (2006).